\documentclass[letterpaper,twocolumn,10pt]{article}

\usepackage{usenix}
\usepackage[nointegrals]{wasysym}  
\usepackage{booktabs}
\usepackage{algpseudocode,algorithm,algorithmicx}
\usepackage{latexsym}      
\usepackage{array,makecell,multirow,graphicx}
\usepackage{amsmath,amssymb, mathtools}
\usepackage{pifont}
\usepackage{latexsym}      
\usepackage{xspace}   
\usepackage{xcolor}   
\usepackage{tabularx} 
\usepackage[skip=8pt, font=normal,labelfont=bf]{caption}
\usepackage{balance}
\usepackage{color,colortbl}
\usepackage{csquotes}
\usepackage{changepage}
\usepackage{url}

\renewcommand{\paragraph}[1]{\vspace{5pt}\noindent\textbf{#1\quad}}
\renewcommand{\subparagraph}[1]{\vspace{5pt}\noindent\textit{#1\quad}}

\renewenvironment{displayquote}{%
   \list{}{%
     \fontsize{9.2pt}{10.7pt}\selectfont
     \leftmargin0.32cm   
     \rightmargin0.32cm
   }
   \item\relax
}
{\endlist}
\newcommand{\authspace}{\hspace{3mm}}

\title{``Why wouldn't someone think of democracy as a target?'': \\Security practices \& challenges of people involved with U.S. political campaigns}

\author{
Sunny Consolvo \authspace
Patrick Gage Kelley \authspace
Tara Matthews \authspace
Kurt Thomas \authspace
Lee Dunn \authspace\vspace{5pt}
Elie Bursztein \\\vspace{5pt}
 \{sconsolvo, patrickgage, taramatthews, kurtthomas, leedunn, elieb\}@google.com\\
Google}

\begin{document}

\maketitle

\begin{abstract}
People who are involved with political campaigns face increased digital security threats from well-funded, sophisticated attackers, especially nation-states. Improving political campaign security is a vital part of protecting democracy. To identify campaign security issues, we conducted qualitative research with 28 participants across the U.S. political spectrum to understand the digital security practices, challenges, and perceptions of people involved in campaigns. A main, overarching finding is that a unique combination of threats, constraints, and work culture lead people involved with political campaigns to use technologies from across platforms and domains in ways that leave them—and democracy—vulnerable to security attacks. Sensitive data was kept in a plethora of personal and work accounts, with ad hoc adoption of strong passwords, two-factor authentication, encryption, and access controls. No individual company, committee, organization, campaign, or academic institution can solve the identified problems on their own. To this end, we provide an initial understanding of this complex problem space and recommendations for how a diverse group of experts can begin working together to improve security for political campaigns.
\end{abstract}
\section{Introduction}

\begin{displayquote}
``What is 100\% true\dots is that foreign adversaries want [campaign] information\dots The faster we all realize that, the better off we're going to be\dots to see politics and campaigns at all levels as a fundamental piece of democracy that needs to be protected\dots. For sure foreign adversaries are trying to attack our systems\dots Why wouldn't someone think of democracy as a target?'' –A participant
\end{displayquote}

Political campaigns, for their role in free and fair elections, are a fundamental part of democracies around the world. Alarmingly, the people and organizations supporting campaigns are under attack. High-profile compromises include John Podesta's personal email in 2016 (during Clinton's U.S. Presidential campaign)~\cite{oremus2016,satter2017}, multiple national party committees in 2015 and 2016~\cite{nakashima2018, bump2018}, Sarah Palin's personal email in 2008 (during McCain's U.S. Presidential campaign)~\cite{zetter2008}, and emails from Emmanuel Macron's campaign (during the 2017 French Presidential race)~\cite{willsher2017}. Lesser known campaigns have also been affected—for example, a candidate whose strategy documents were leaked in 2016 lost their primary race for a seat in the House of Representatives~\cite{liptonshane2016}. These examples illustrate how security attacks on people involved with campaigns can damage campaigns, potentially change election outcomes, and undermine public trust. Thus, digital security is an important part of winning elections and is a nonpartisan imperative for protecting democratic institutions.

Attackers may target anyone who is affiliated with campaigns, including candidates, campaign staff, political consultants, committee staff, and more. These people face an outsized security risk compared to the general population. Successful attacks can also carry disproportionate impacts. A growing literature on at-risk user populations (e.g.,~\cite{mcgregor_investigating_2015,tadic_ict_2016,matthews_stories_2017,guberek_keeping_nodate,vitak_i_2018}) demonstrates the importance of understanding the different and complex kinds of risks these users face. Because of the elevated risk and outsized harms facing people involved with campaigns, it is important that the security community understand their unique perspectives and needs.

Despite the critical role that campaigns play in democratic elections, there is little in the literature about the security perceptions and needs of the people who are involved with campaigns—that is, those who work on, with, or in support of them. Prior work on election security has focused largely on securing election infrastructure~\cite{natlacademy_securing_nodate,belfer_state_nodate,mcfaul_securing_2019,alvarez_securing_2020}, ensuring election outcomes are accurate~\cite{lindeman_gentle_2012,hall_implementing_2009,alvarez_securing_2020},
investigating the nature and impacts of mis/disinformation operations on democracy~\cite{schia_hacking_2020,bradshaw_global_2018,badawy_analyzing_2018,morgan_fake_2018,faris_partisanship_2017,pope_cyber-securing_2018},
and describing politically-motivated trolling operations and their impacts on citizen participation ~\cite{aro_cyberspace_2016,bradshaw_troops_2017,forestal_architecture_2017}. Several organizations have produced guidance for campaign workers, embedding knowledge of the security protections they should employ~\cite{belfer_cybersecurity_2017, lord_device_2019,fbi_protected_nodate,noauthor_election_2018,wallach_howto_2017,ddc_five_nodate,noauthor_usc_nodate,zeltser_cybersecurity_2018}.
However, these studies and guides leave a gap in the security community's understanding of how people involved with campaigns approach security technologies and the related barriers to adoption these users experience.

In this work, we study the security perceptions, practices, and challenges of people involved with political campaigns through qualitative research conducted with 28 participants across the U.S. political spectrum. Our participants represented a wide range of roles and organizations affiliated with campaigns, from local to Presidential elections. Our study explores the following research questions:

\begin{itemize}
\item \textbf{RQ1:} What threats do people involved with campaigns perceive that they face? What outcomes do they believe could result from the threats?
\item \textbf{RQ2:} How do people involved with campaigns use technology in their day to day work? What digital security vulnerabilities does this introduce?
\item \textbf{RQ3:} How well do people involved with campaigns understand or adopt security best practices for campaigns?
\item \textbf{RQ4:} What work culture or contextual factors influence the adoption of security best practices within campaigns?
\item \textbf{RQ5:} How do people involved with campaigns think security practices or technology might be improved?
\end{itemize}

The key contribution of this paper is an initial understanding of campaign security from people involved with them, and recommendations for how to begin to improve this complex problem space from a user-centered perspective. We find that a unique combination of threats, constraints, and work culture lead people involved with political campaigns to use technologies from across platforms and domains in ways that leave them—and democracy—vulnerable to security attacks. Security is a relatively new concern in the campaign space, and parts of this well-established sector have yet to adapt. In particular, this population works in a fast-paced, hectic, temporary environment, where winning the election is the top priority and anything that does not clearly contribute to that is perceived to be a waste of time. This population also tends to not have formal technology training, and their digital security knowledge is limited. This results in a work environment in which particularly sensitive data—communications and files—are vulnerable due to weak security practices, including the ad hoc use of strong passwords, two-factor authentication (2FA), encryption, and access control restrictions.

We detail campaign work culture and practices so that technology creators, together with a diverse group of experts who are trying to support this population, can understand barriers campaigns are likely to face when trying to adopt strong security practices. We suggest that the longer-term goal of those who are trying to support this population should be to shift the work culture on campaigns to prioritize security, and that in the near-term, it should seek further consensus around the security actions campaigns should prioritize to inform security education; investigate how to coordinate the standardization of account security protections (including 2FA, password managers, and multi-tenant accounts); and improve the affordability of security technologies and training. Progress on these issues are initial steps toward helping to ensure that democratic elections focus on candidates' messages and merits, not on their digital security practices.
\section{Related Work}

We frame our study on the security practices of people involved with political campaigns in the broader context of election security research, existing guidance for campaigns, and related studies of at-risk populations and high-risk work environments. 

\subsection{Digital threats facing democracy}

Our research is part of a much broader area exploring how to secure democracy against digital threats. Collectively, these discussions of “security” incorporate a broad view of how digital information or systems may be used to harm the democratic process or relevant institutions. For example, Whyte~\cite{whyte_cyber_2020} describes “cyber-enabled information warfare” as attempts to influence democratic politics with tactics such as hacking private information sources within a democracy, then using what was stolen or other messaging to threaten or undermine the credibility of the democracy's sources of information. Looking broadly across various observed election and political interference operations, Herpig et al. ~\cite{herpig_securing_2018} categorized digital attack tactics against democracy to include: manipulating data (e.g., votes), eroding trust in democratic processes, denying legitimate actors access to critical data or infrastructure, espionage, leaking data, persuading voters, and blackmail. Herpig et al. further described these tactics as seeking to undermine democracy from different angles, such as changing the outcome of elections, delegitimizing democratic processes, harming reputations, or creating barriers to government operations or relations.

Within this broad body of work on securing democracy from digital attackers, researchers have focused on the security of elections~\cite{natlacademy_securing_nodate,belfer_state_nodate,mcfaul_securing_2019,alvarez_securing_2020}, information campaigns intended to misinform and polarize the public or targeted groups~\cite{schia_hacking_2020,bradshaw_global_2018,badawy_analyzing_2018,morgan_fake_2018,faris_partisanship_2017,pope_cyber-securing_2018,tenove_digital_2018}, and trolling operations aimed at suppressing citizen participation in democracy~\cite{aro_cyberspace_2016,bradshaw_troops_2017,forestal_architecture_2017,tenove_digital_2018}.

Our research distinctly focuses on the security practices and challenges of people who are involved with political campaigns, which tend to operate in the months leading up to and weeks just after election day. 

\subsection{Protecting digital security in elections}

The bulk of the literature on election security focuses on election infrastructure and the verifiability of election outcomes. Researchers have broadly explored the security of election infrastructure in the U.S. (e.g.,~\cite{natlacademy_securing_nodate,lazarus_applying_2011,belfer_cybersecurity_2017,mcfaul_securing_2019,alvarez_securing_2020}) and other countries (e.g.,~\cite{haines_how_2020,springall_security_2014,brown_cybersecurity_2020}). Evaluations of specific voting machines (e.g.,~\cite{feldman_security_2007,appel_new_2009,blaze2017}), ballot marking devices (e.g.,~\cite{bernhard_can_2020}), and voting applications (e.g.,~\cite{specter_ballot_2020,specter2021}), as well as post-election audits (e.g.,~\cite{lindeman_gentle_2012,hall_implementing_2009,alvarez_evaluating_2012}), aimed to identify insecurities in voting systems and ensure votes were counted correctly. Bernhard et al.~\cite{bernhard2017public} outlined the requirements of secure, verifiable elections, and reviewed current electronic voting systems and auditing approaches to highlight open issues. To help U.S. state and local election officials understand and protect against security threats to elections, The Belfer Center for Science and International Affairs' Defending Digital Democracy Project (D3P) published a security guide for them~\cite{belfer_cybersecurity_2017}. However, election infrastructure and the people managing it are often separate from those involved with political campaigns, who we studied.

The vast majority of what is written about the security of people involved with campaigns can be found outside of peer-reviewed literature. The press and bloggers have anecdotally described some security practices of campaign workers and adoption barriers they face, such as certain work culture issues, use of personal accounts, inconsistent use of encrypted messaging, under-use of 2FA, and challenges setting up security keys and password managers~\cite{ceglowski_what_2019,volz_2020_2019,cramer_democrats_nodate,marks2019}. Though little is published about how campaign workers approach security, we do know they are at risk. The U.S. government~\cite{evanina2020}, technology companies~\cite{huntley_how_2020,burt_new_2020} and the press~\cite{perlroth2019, matishak2018, volz_2020_2019} have reported on security threats campaigns face, focusing on nation-state attackers utilizing phishing and disinformation. We seek to illuminate the perspectives and practices of campaign workers in a rigorous, qualitative study, to help address these serious threats to democracy.

\subsection{Efforts to support campaign security}

Several groups have developed security guides, checklists, or playbooks for people on campaigns. These guides were often created by people who have worked on or with campaigns, and embed knowledge of what security protections campaigns need most. The Belfer Center's D3P has produced several guides for campaigns, including a five-item prioritized checklist of security advice~\cite{belfer_cybersecurity_2017}. Bob Lord, Chief Security Officer of the Democratic National Committee (DNC), published a device and account security checklist aimed at people working with campaigns~\cite{lord_device_2019}. Other guides focusing on campaign security include those from the FBI's Protected Voices initiative~\cite{fbi_protected_nodate}, the Center for Democracy \& Technology~\cite{noauthor_election_2018}, Professor Dan Wallach of Rice University~\cite{wallach_howto_2017}, Defending Digital Campaigns (DDC)~\cite{ddc_five_nodate}, the USC Election Cybersecurity Initiative~\cite{noauthor_usc_nodate}, and Lenny Zeltser of Axonius~\cite{zeltser_cybersecurity_2018}. Collectively, these guides prioritize use of secure cloud providers, strong account security (2FA, strong passwords, password managers), encryption for communications and stored data, and security training. But across the set of guides, a plethora of other actions are recommended, with extensive technical detail and differing priorities, which may overwhelm campaign workers who have limited time and security knowledge.

There is limited organizational support to help campaigns wade through disparate, complex guidance. Federal Election Commission (FEC) campaign finance rules require that campaigns use their own money to purchase security products or services. In only the past two years, the DDC\footnote{\url{https://www.defendcampaigns.org/}}, a non-profit organization, received special permission from the FEC to provide low- and no-cost security products and services to committees and campaigns meeting certain criteria, across parties~\cite{fec2018}. The DDC has partnered with a host of companies to begin implementing key aspects of these guidelines on federal campaigns. But many campaigns may not be able to receive security support via these specific FEC exemptions.

Overall, political campaigns have limited, fledgling support for security, while threats from adversaries rapidly grow and change. This study of campaign workers' perspectives and practices is an important early step toward ensuring new and modified technologies will meet user needs. 

\subsection{At-risk populations \& security}

Our study joins a growing body of research on at-risk populations—technology users who are specifically targeted for security or privacy invasions because of who they are (e.g., transgender individuals ~\cite{lerner_privacy_2020,scheuerman_safe_2018}, people with visual impairments~\cite{hayes_cooperative_2019,ahmed2015privacy,akter2020}, and people with learning disabilities~\cite{marne_learning_2017}), what they do (e.g., journalists~\cite{mcgregor_investigating_2015,mcGregor2016} and activists~\cite{tadic_ict_2016}), where they are (e.g., people in South Asia~\cite{sambasivan_toward_2019,ahmed_everyone_2019}), or who they are/were with (e.g., survivors of intimate partner abuse~\cite{freed_digital_2017,matthews_stories_2017,tseng_tools_2020,havron_clinical_2019}, survivors of trafficking~\cite{chen_computer_2019}, and people targeted by other close relations~\cite{levy_privacy_2020}). Some at-risk populations may be susceptible to attacks because they don't know how to avoid them or lack the resources to recover (e.g., children~\cite{jeong_lime_2020,zhao_i_2019}, older adults~\cite{frik_privacy_2019,mcneill_privacy_2017,mentis_upside_2019}, undocumented immigrants~\cite{guberek_keeping_nodate}, refugees~\cite{simko_computer_2018}, and people experiencing financial insecurity~\cite{vitak_i_2018,sleeper_tough_2019}). Groups may be uniquely at risk at the intersections of any of the above characteristics (e.g., low-income African Americans in New York City~\cite{elliott_straight_nodate}).

Despite this growing literature, the vast majority of research on user perceptions of digital security focuses on general user populations (e.g.,~\cite{wash_folk_2010, kang_my_2015, herley2009, whitten1999}). But technology designed wholly based on the feedback of general users may not adequately address the threats to at-risk populations. An important contribution of this and other work with at-risk users is to add nuance to the security community's understanding of the different and complex kinds of risk some users face that do not necessarily fit into general models of usable security.

\subsection{Security in high-risk work environments}

Our work on political campaign security shares similarities with prior research on workplace security. Many types of organizations must balance investments of time and money on security with other priorities, and those priorities depend on many factors, such as assessed risks, business or work practice needs, organizational maturity, and more \cite{mcGregor2016, chen2016, alahmari2020, huaman2021}. Prior work on high-risk work environments shows the value in understanding unique working conditions that both elevate risk and augment threat models. Examples include journalists and news organizations \cite{mcgregor_investigating_2015, mcGregor2016} and small businesses \cite{alahmari2020, chen2016, huaman2021}.

McGregor et al. \cite{mcgregor_investigating_2015,mcGregor2016} studied journalists and their organizations. Similar to our findings for campaign workers, journalists were most worried about their communications being targeted by sophisticated attackers, but were not widely adopting security technologies \cite{mcgregor_investigating_2015}. Also similarly, journalists noted having limited time and technical expertise available to implement security protections. Unlike campaign workers, many (though not all) journalists worked in stable news organizations, with at least some IT staff and security infrastructure and policies that emphasized account security and anti-phishing \cite{mcGregor2016}. Individual journalists were more concerned about surveillance from their own government (than from foreign governments, as we found for campaign workers), making encryption an important perceived protection. Unique to their work practices, journalists and their organizations often focused on protecting sources, who were typically the ultimate target of security attacks, but neither had the ability to control security protections that sources used.

Small businesses are similar to campaigns in that they often lack the financial resources and expertise to implement security protections and underestimate security threats, despite being targeted by digital attackers who perceive them to be soft targets \cite{alahmari2020,chen2016}. However, recent work by Huaman et al. \cite{huaman2021} suggests that many small businesses are now implementing a range of technical security measures and have a basic awareness of security issues, though adoption of organizational security measures lags behind. Further, older small businesses studied in \cite{huaman2021} were more likely to have certain technical security measures in place (like antivirus software, firewalls, and a backup strategy). Unlike short-lived, amorphous campaigns, investments in security have a different return on investment and may be easier to prioritize, especially as a company matures.

As we will show, campaigns are unique compared to these previously studied workplaces due to their extreme transience, amorphous boundaries and leadership structure. Campaigns also face sophisticated, nation-state attackers who often aim to damage democracy, not just the campaign. At the same time, the fast pace of campaigns and fixed duration of an election cycle have historically instilled an attitude amongst people involved in campaigns that effective security protections are beyond practical reach. These qualities create a novel threat landscape for the security community to understand.


\section{Methodology}
We conducted semi-structured interviews with 28 participants who had experience working on, with, or in support of political campaigns in the U.S. Here we describe our participants, recruiting, procedures, data collected, analysis, ethical considerations, anonymization, and study limitations.

\subsection{Participants \& recruiting}

We recruited people who were or had been involved with U.S. political campaigns, aiming for breadth in political party, role, organization, and level of experience. We identified potential participants who represented this population and directly contacted them with help from known contacts or via their social media profiles. We ensured representation of a variety of roles and organizations. Our participants included campaign staff (17 participants), support organization executives/staff (12), consultants (10), party committee executives/staff (10), IT/security support (6), and candidates (4). They had collectively worked directly on or with a campaign (22); with a support organization such as a super PAC, think tank, university, or software provider (12); or on a national/state party committee (10). Relative to their role(s) with campaigns, 17 were from the Democratic party, 10 Republican, and 4 Non/Bipartisan (3 had served non/bipartisan as well as partisan roles). Twenty-five had been involved with Federal campaigns, 8 with State, and 9 Local. All participants named multiple roles they had held working on or with campaigns, hence numbers will not total to 28.

Though we recruited people with a range of experience, we focused recruiting on those with extensive, long term experience, as they could speak to changes in campaign culture and practices over time. We also placed an emphasis on recruiting people who had experience with campaigns following the high profile attacks of the 2016 election cycle ~\cite{oremus2016,satter2017,nakashima2018,bump2018}. Seven participants reported 20+ years of experience in politics, 13 reported 10-20 years of experience, 1 reported less than 10 years, and 7 did not clearly specify. 

We conducted interviews from May 2019 to March 2020, prior to COVID-19 shutdowns in the U.S. All participants had a chance to review consent materials prior to the interview starting. No financial incentive to participate was provided.

\subsection{Procedures \& data collected}

The lead interviewer began the interview session by going over with participants a two-paged consent document. We emphasized their agency to end the interview or pause the audio recording at any time. All participants gave informed consent. Next, our first interview question asked about the participant's background working with political campaigns. Then we drew from the following interview topics, in a semi-structured way: 

\begin{itemize}
\item \textbf{Threats:} Security threats participants believed campaign(s) or related organizations faced (probing about specific attacks, attackers, what data was at risk, and what harms could result).

\item \textbf{Vulnerabilities:} Areas of security they felt could be stronger in campaigns or political organizations (probing about the various security practices that the participant brought up, best practices in the participant's opinion, and perceived barriers to adoption).

\item \textbf{Work \& technology practices:} Communication and account practices of the participant and their colleagues, and any security training they had taken. This included their understanding and use of 2FA.

\item \textbf{Long-term outcomes:} The impact they felt the current campaign security landscape could have on the future.
\end{itemize}

The same two members of the core research team attended the interview sessions to either lead the interview, or take notes plus ask occasional follow up questions (except two sessions, which were attended by only one of the researchers). One researcher led 19 sessions; the other led 5. Most interview sessions (N=24) included one participant; two sessions included two participants; and one session included three participants. The three sessions with more than one participant were arranged at the request of our participants. In those three sessions, all participants were encouraged to respond to all questions asked. These participants served different roles (e.g., candidate and campaign manager), that allowed them to offer different perspectives and add context to responses. To ensure security and confidentiality, we conducted all but three of the interviews entirely in person; the three interviews that included a remote component employed video or phone conferencing. We audio recorded sessions with participant consent. See the Appendix for our interview script, recruiting material, and consent form.

Sessions lasted 42-118 minutes ($\mu$=75). We collected 27 hours of audio recordings in total (one participant declined recording, but detailed notes were taken and analyzed with permission). Two members of the core research team transcribed the audio in a verbatim style (i.e., a word-for-word transcription, including fillers and non-verbal utterances), producing 948 pages of data. For the safety, security, and privacy of participants, only members of the core research team transcribed the recorded audio and had access to the transcripts.

\subsection{Analysis}

We used an inductive~\cite{thomas_general_2006}, thematic analysis~\cite{braun_using_2006} to analyze the transcript data. The inductive (or data-driven) method was appropriate because campaign security issues, from a technology user’s perspective, is a novel empirical topic with limited prior work upon which to base a deductive approach. We chose a thematic analysis for its flexibility and because we believed the output (a set of patterns coupled with rich detail) would be most useful for an audience of technology creators, building an early understanding of this population.

To begin, three members of the core research team (referred to as “we” from now on; these three were made up of the two researchers who conducted all of the interviews, and the two who transcribed all of the audio) deeply engaged with the interview data by listening to audio recordings and reading the transcripts multiple times. We used these activities to memo and summarize all the data, and to brainstorm an early list of inductive codes. One of us, an expert coder, then coded all the data, inductively building and revising the list of codes as the coding progressed, to ensure it described the full data set. To reach agreement on codes, we met regularly throughout the coding process to develop, discuss, and refine codes. We did not set up the coding process to use inter-rater reliability measures (IRR) to assess agreement for several reasons covered in McDonald et al.~\cite{mcdonald_reliability_2019}: a set of themes was the goal of our analysis, not a set of codes (we discuss stakeholder checks below to establish the credibility of these themes); and our data was complex and nuanced, and the need to meet IRR metrics can cause complex concepts and nuance to be lost. See the Appendix for the list of codes pertaining to the themes in this paper.

Next, in regular meetings and independently, we searched for themes in the coded data, using and iteratively adding to/revising the memos, summaries, and coding. We then reviewed our ideas for themes by revisiting the relevantly coded data, further memoing and revising themes, codes, and coding of data as needed. We also created rough thematic maps of our data, to aid discussions. In this paper, we focus on themes in our dataset that we determined were most important (in terms of mitigating risk to the population) and pervasive (as reported by the set of participants).

We performed several credibility checks on our findings~\cite{thomas_general_2006} to ensure they were accurate and covered the most salient themes about campaign security. First, we summarized initial findings and discussed them with domain experts who had experience on or with campaigns. Second, we shared our summarized findings in a report and presentation, and collected feedback from a multidisciplinary group of experts who participated in a roundtable that we ran in Summer 2020. Forty-four experts attended the roundtable, not including the event organizers; 13 attendees were from the Republican party, 12 from the Democratic party, 10 from tech companies, and 9 from academia or non-partisan political nonprofits (see \cite{roundtable-report} for more about the roundtable and the 28 organizations that were represented). Finally, we talked with security experts to ensure we accurately portrayed the security attacks and technologies in our findings.

\subsection{Ethical considerations}

When working with any at-risk population, ethical considerations are essential to the study design, analysis, and reporting. Since campaigns face targeted threats from sophisticated attackers, our focus was on ensuring that participation in this study would not harm participants and the population more generally. We conducted the vast majority of interviews in person, to minimize the risk of surveillance. We protected the data produced during our study—including audio, memos, and transcripts—by encrypting all records at-rest, restricting access to only the core research team (and institutional administrators), and requiring 2FA with a physical security key to access the information. 

Anonymity required extra attention. Some of our participants were concerned about their anonymity, having experienced harassment and other attacks as part of their roles in politics. We also considered the risk of re-identification in light of their highly public careers. Our goal was to protect all participants' anonymity by using a conservative reporting standard. Thus, when reporting our results, we specifically omit unique details, phrases, or words in quotes from participants to mitigate de-anonymization, and likewise report only coarse, aggregate demographic features. We do not attribute quotes, even using pseudonyms, to mitigate the collection of information about any one participant. Instead, we report that the 32 quotes in this paper are from 18 out of 28 participants, and we did not quote any one participant more than 4 times.

To ensure our work did not put participants or others who work with campaigns at risk, our study plan, data handling practices, and final study report were reviewed by a set of experts from our institution in domains including ethics, human subjects research, policy, legal, security, privacy, and anti-abuse. We note that our institution does not require IRB approval, though we adhere to similarly strict standards.

\subsection{Limitations}

Our study was limited to participants in the U.S. where campaign operations, security advice, and threat perceptions may differ compared to other regions. Our small sample favored in-depth reporting over broad generalizability across all campaigns and roles. Even though our participants had experience working on city or regional races, their experiences and feedback skewed toward state-wide or federal races. They also skewed toward professionals with more years of experience. We note that comparing across parties was not a goal of this work, and while we did not observe differences, we cannot rule out that differences could exist. All of that said, we reached data saturation with our interviews, that is, a point at which we found no new themes as interviews progressed, indicating that the themes presented in this paper are robust for the population represented by our participants. Finally, this study was affected by the standard limitations of self-reported data, including recall bias and observer bias.
\section{Results}
Our results point to three main security challenges facing campaigns: (1) a work culture where security is not emphasized or prioritized; (2) ad hoc security practices used by people involved with campaigns, across many technology platforms, to manage data that attackers often seek; and (3) a nascent, though growing, understanding of the threats facing campaigns that guides decision making. While every campaign may face a different subset of these challenges, the themes we highlight were raised by multiple participants. See Table 1 for a preview of these themes.

\begin{table*}[ht]
    \centering
    \begin{tabular}{p{3.8cm} p{13.1cm}}
         \textit{Theme} & \textit{Example Quote} \\ \midrule
         
\textbf{Work culture: Busy} &
``It's pretty hard to overstate the chaos of most campaigns. And so, there's risk everywhere.'' \\ \midrule

\textbf{Work culture: Security is not a priority} &
``You have to change the culture of the <political campaign> industry, so that it's expected rather than an afterthought. And right now, I think it's mostly an afterthought.'' \\ \midrule

\textbf{Tech practices: Communications are sensitive} &
``The `e' in email stands for evidence.'' \\ \midrule

\textbf{Tech practices: Account security is weak} &
``A <Provider> account with no two-factor and a password of 1234 is very much a thing… So, the challenge is that you have many different accounts—both the campaign accounts, and perhaps even more importantly, the personal accounts that people use.'' \\ \midrule

\textbf{Perceived threats: Primarily nation-states} &
``Ever since I've worked here, <we> have been worried about a foreign entity hacking us… I've never worried about <the other party> hacking us.'' \\ \midrule

\textbf{Perceived threats: Harm to democracy} &
``<Security attacks> changed the outcome of the election in 2016… It erodes our democracy, our institutions, it erodes confidence and trust… That's why <nation-states> are looking to interfere, because they had such success in really shaping an election in... the biggest democracy in the history of the world.'' \\ \bottomrule
         
    \end{tabular}
    \caption{Select themes and example quotes from results. Some of our main study themes, chosen for their importance (in terms of mitigating risk to the population) and pervasiveness (as reported across participants), are listed here with representative quotes.}
    \label{tab:themes}
\end{table*}

\subsection{Campaign work culture}
We detail how campaign work culture, influenced by many constraints, created a challenging environment for security.

\paragraph{Winning.}
Campaigns exist to win elections—participants emphasized it as the top priority. For decades, voter and donor outreach tasks have been considered among the most important to achieving this goal.\footnote{Note that campaigns prioritize a host of tasks beyond outreach, and it is not the purpose of this work to analyze priorities across all tasks. Instead, we focus on understanding where security fits in the prioritization, and highlight certain higher priority tasks that contribute to security vulnerabilities.} This meant communicating with voters and donors wherever they were, including at their doorstep, on TV, in email, and on various social media platforms. Only recently has digital security become an important need for campaigns, and prioritizing it required trade-offs—for example, spending less time and money on voter and donor outreach to free up time and money for new digital security needs. This was described as a hard sell to decision makers on campaigns (these varied, but could include the campaign manager, senior staff/consultants, or the candidate).

\begin{displayquote}
``A \$25-million Senate race is pretty standard these days. And the committees want you to spend all that money on voter contact and direct communication with voters, which is smart, but there isn't the sort of <security> infrastructure setup that helps someone who has never run before, who has no idea that <nation-state> operatives can be trying to break into a Senate race and influence it.'' –A participant
\end{displayquote}

\paragraph{Transience.}
Participants emphasized that campaigns were transient—that is, they needed to fully ramp up on a short timeline (ranging from about 3 months to a little over a year, depending on the type of race), with a clear, immovable end date (election day). Staff were only hired for the duration of the campaign, and frequent job changes were common in the political sector. For technology, transience meant there was not time to develop much IT/security infrastructure or institutional knowledge before a campaign shut down. To cope, dormant accounts and infrastructure from a previous campaign might be reused, entirely new infrastructure rapidly brought online, or a combination of the two.

\begin{displayquote}
``Most campaigns basically start 6 months before the election. Maybe 8 months… So what happens is that you quickly hire a bunch of people who work for you for maybe a hundred days, and then disappear… And so, HR? Forget it. IT security standards? Forget it. Two-factor authentication? Forget it... There's no ongoing, continuing infrastructure. There's no culture… So I think one challenge you have with campaigns is they're totally transient... There are very few incentives for any kind of <security> rigor. Because you're up against the clock, and faced with the ticking clock, everything pales.'' –A participant
\end{displayquote}

\paragraph{Busy workers \& tight budgets.}
People on campaigns were described as chaotically busy. Tight budgets and a short timeline meant limited staff working at break-neck pace. Money was unlikely to be allocated for security expertise or technologies, especially in the beginning of a campaign as habits and processes were being developed. Busy people had less time to enact unfamiliar security measures.

\begin{displayquote}
``There's the wolf in your yard, the wolf on your deck, the wolf in your house, and the wolf in your pants. Cybersecurity is like the wolf in your yard. You can't even get there because you have these day-to-day things that consume your life. It's very difficult to look past that wolf in your pants… The focus is not, ‘What if <nation-state> hacked us?' We'll deal with that when it happens.'' –A participant 
\end{displayquote}

The themes in this paper apply to campaigns at all levels, but differing budgets impacted campaigns' ability to mitigate security problems. Larger campaigns, like Presidential or contested races that gained national attention, tended to have more resources to apply to security if they chose to. Participants reported that larger campaigns were more likely to hire security expertise and buy managed accounts for staff. Senate and Gubernatorial races were often medium sized and fairly well-resourced, but down-ballot from these, campaigns were described as typically very small (possibly only hiring one or a few staff members) and extremely resource-constrained. Regardless of campaign size, participants perceived that security investments were usually not prioritized.

\begin{displayquote}
``You spend X dollars on this, translates to X votes. Security is not baked into that... So every dollar we spend <on security>, is a dollar taken away from a vote that they're going to <use to> buy a commercial in <City>.'' –A participant
\end{displayquote}

\paragraph{Amorphous boundaries.}
Campaigns were described as having amorphous boundaries in terms of life cycle, people, and data access. Regarding life cycle, the start of a campaign was described as a gray area. Conversations about whether to run for office, fundraising, and research tended to occur before a campaign became official. This meant using accounts outside the campaign’s domain (which likely had not yet been set up). When a campaign did become official, it was described as common for many people to be onboarded and accounts to be set up before security was considered.

Regarding people, transience and tight budgets were described as motivators for many professionals to be more permanently employed by consulting firms that supported multiple campaigns, with their domain accounts and security practices not controlled by any one campaign. Consultancies were described as ranging in their security standards: some consultancies had developed security standards that largely followed best practices, which they sometimes tried to apply on campaigns; some consultants did not and relied upon basic consumer technologies for work use. How rigorously they applied security protections seemed to depend on their leadership and knowledge about security. Furthermore, tight budgets meant campaigns relied on the free labor of hundreds (or even thousands) of volunteers. These amorphous boundaries complicated information access decisions and security (e.g., who gets access to what data? who gets an official campaign account? who must follow security standards, and how can they be enforced?), in an environment where security knowledge to cope with such complexity was often limited.

\begin{displayquote}
[Where is a campaign’s sensitive data kept?] ``There are a lot of third parties in this environment. You're going to have your media consultant, your finance consultant, your <third party vendor> to process donations... I don't know the degree to which people are running reports and downloading them to their own computers to manipulate the data.... This is a very squishy boundary operation. So I think <sensitive data is> all over the place'' –A participant
\end{displayquote}

\paragraph{Security knowledge.}
Security and IT knowledge was generally lacking on campaigns, though this varied with the campaign's size and funding. Trained security and IT professionals tended to have more job security and financial opportunity in industry, where organizations were not as transient or budget constrained, and where a greater emphasis was often placed on security. Even if security or IT professionals were available to hire, spending money on security or IT was perceived as cutting directly into traditionally higher priority tasks (like voter and donor outreach). This lack of security and IT expertise meant that staffers with no (or outdated) technology training were often responsible for the campaign's technology infrastructure and practices. For campaigns that hired someone to handle IT or security (and again, these tended to be larger races), it was typically a vendor. Vendors usually did not work full-time for the campaign, and sometimes they set up technology and were then ``done.'' They also tended to be brought in once campaigns were reasonably established and work practices were already formed and harder to change. 

Furthermore, our data included many stories of people on campaigns who did not understand or know how to effectively use the security technologies that were recommended to them. Indicative of their limited security knowledge, our participants described colleagues who did not believe they were targeted any more than the general population, and a smaller number of our own participants found it difficult to believe they may be targeted.

\begin{displayquote}
``It steps into the realm of paranoia, conspiracy theories, and self-aggrandizing if you think that your communications are being targeted. And so, despite the fact that a lot of people in politics have big egos, they still wouldn't assume <they are targets>.'' –A participant
\end{displayquote}

This lack of concern contributed to security not being prioritized. Participants described awareness on campaigns as starting to shift: the high profile attacks during the 2016 election cycle~\cite{oremus2016,satter2017,nakashima2018,bump2018} were noted by nearly all participants as a catalyst for increased security concern, though not yet for widespread adoption of protective security behaviors. And in the cases when security actions were taken, the lack of security knowledge on campaigns contributed to vulnerabilities in security configurations (as we will cover later in the results).

\begin{displayquote}
``My experience with political campaigns is they were broadly ignorant and negligent at pretty much all security measures until John Podesta's email showed up. And then suddenly, everybody got worried... But there wasn't a significant increase in security measures on a lot of these campaigns... a simple example being two-factor authentication on email... People are not using it.'' –A participant
\end{displayquote}

\paragraph{Summary: Security is not prioritized.}
All of the characteristics above contributed to a work culture of campaigns in which security was seldom prioritized. As reported by participants about the campaigns they had worked with: digital security—a relatively new need for campaigns—was not commonly viewed as essential to winning elections. Other traditional tasks were seen as more important and thus prioritized, often at the expense of security. Many people on campaigns did not understand the threats they faced, and thus lacked the motivation to spend precious time and effort minimizing risk. And even if they were concerned, people on campaigns felt too busy to spend much time on security, especially since many were unfamiliar with how to effectively use security technologies. Finally, the amorphous boundaries of campaigns significantly complicated (already hard to justify) security efforts by requiring more people to coordinate to protect data that was spread across domains and accounts, starting even before a campaign was official.

\subsection{Campaign tech practices \& vulnerabilities}
People involved with campaigns depended on a variety of accounts and technology platforms for communications and file storage. We found that disparate security practices for managing these sensitive resources—such as encryption, authentication, and access control—compounded the threats they faced. We describe issues with how people on campaigns handled security compared to current best practices for this population, and highlight the vulnerabilities these gaps introduced.  We note that this section focuses on participants’ \textit{perceptions}.

\subsubsection{Sensitive data: communications \& files}
Most (although not all) of our participants believed or cited evidence that their communications—work and/or personal—had been targeted by attackers. Participants explained why: candidates and their campaigns rely on reputation, and almost any communication by someone involved with their campaign could be used to cause damage. Even personal communications, unrelated to the campaign and housed in consumer accounts, could create a problem for a campaign if leaked (Pizzagate~\cite{robb_pizzagate_2017} is a well-known example). Participants also described files as being potential targets (especially those containing strategy, voter, or donor information), which tended to be stored in cloud accounts or as email attachments.

\begin{displayquote}
``Let's say the candidate is just complaining about something. Or the staff is just complaining about something. Someone could very easily grab that and turn it into a thing, when it was actually just a ‘human moment.' And now you've derailed the whole campaign. You've taken it off message. You're talking about whatever stupid thing was in that email, rather than whatever you really want to be talking about.'' -A participant
\end{displayquote}

\paragraph{Ad hoc storage, sharing, \& provider usage for data.}
People involved with campaigns stored data across many domains and providers, including work and personal email, social media accounts, and chat accounts. Traditional workplace access control policies—such as restricting communication and file access to a campaign's domain or workplace accounts—were not typically employed by the campaigns our participants had worked with, given their need to share access with external committee staff, consultants, vendors, volunteers, or even a candidate's family. A campaign's quick ramp up often meant no pre-existing IT infrastructure or policies, so early campaign staff and consultants' preferences often drove the decisions of what technologies to use. And since these individuals continued to work in politics, they sometimes kept important communications and files in their own accounts that would persist beyond a single campaign. Preferences were also driven by practical time constraints—some participants perceived that there was not enough time to set up and learn new technologies for each campaign.

The security configurations of these various accounts were often not controlled by (or sometimes even known to) the campaign. These dynamics also tended to result in what participants described as an ``ad hoc'' set of communication practices, for example, communicating in whatever technologies people already had set up, were convenient on the devices they already used, already housed the necessary contact information, etc. Participants described the act of coping with these complex, cross-provider technology setups as involving inherent gray areas—what technology should be used to communicate various content among the many options?; who should be included on emails or file access control lists (ACLs) when boundaries were amorphous?; who should be removed from file ACLs and how should that be accomplished when the campaign ends abruptly?

\begin{displayquote}
``Imagine an email on a sensitive topic. It's going to consist of maybe 5 or 6 people: a couple of them are going to be on the campaign email account, a couple of them on <consumer accounts>, somebody is going to have their own domain, but it's not like there's any IT behind it... These are not communications that are happening within a single organization that can be locked down. And they're not happening for the most part with people or organizations that are very sophisticated about security.'' –A participant
\end{displayquote}

\paragraph{Practices protecting communications \& files.}
Relevant to the challenges above, participants talked about protective practices that our findings suggest are on the rise. Most prominently among these include encrypted communications and secure data sharing in the cloud. Even for campaigns and related organizations that had adopted these tools and practices, participants described usability issues and inconsistent use across people and contexts as problems.

\subparagraph{Encrypted communications \& files.} Participants described that many campaigns and related organizations were adopting encrypted communications tools, with Signal and Wickr being mentioned most often. But decisions about when to use encrypted communications over standard email or chat were often still ad hoc. Encrypted file storage and sharing systems were less commonly described, and usability was noted as a barrier. Participants thought encrypted tools were harder to use than other tools, like standard email, messaging, or file storage/sharing.

\begin{displayquote}
``Encryption, having all your communications be encrypted so that they aren't vulnerable to a hack, I don't think is totally figured out yet. People use <encrypted chat tool>... Sensitive communication that you might typically do on email, you move over to some kind of encrypted system. I think it's pretty ad hoc still... How do you protect your documents so that they're not available to a system hack, but also make them usable?... We looked into how to encrypt documents… We couldn't do it because the bar and the barrier to entry for the people that work at <organization> every day to get around the encryption... was too cumbersome.'' –A participant
\end{displayquote}

\subparagraph{Secure data sharing in the cloud.} Participants described nearly pervasive use of the cloud to store and share sensitive campaign data, like email and files. Fewer participants—and primarily those from consulting firms, committees, and organizations—discussed using best practices for cloud systems, like auditing file ACLs, establishing secure sharing norms, or employing retention policies. Importantly, moving sensitive data to the cloud made strong account security essential, and as we will cover in the next section, this was often lacking.

However, we heard multiple stories of oversharing mistakes with cloud files and email. Some of these cases were due to usability: default file sharing settings that gave access to anyone who had the link sometimes resulted in sensitive campaign files being shared more broadly than expected.

\subsubsection{Accounts \& Authentication}
Participants described account security practices on campaigns that introduced vulnerabilities, including the use of personal accounts for campaign data, account sharing, transient account ownership, under-use of strong 2FA, and weak password practices. These weaknesses were exacerbated by the many accounts that campaign workers used, and security was complicated by the cross-platform nature and campaigns' inability to enforce policies for accounts they did not control.

\paragraph{Many targeted accounts.}
Participants described an environment in which candidates, campaign staff, and consultants used numerous email, chat, social media, and cloud accounts, any of which might be targeted by attackers. While not directly prompted, participants named commonly used providers like Apple, Dropbox, Facebook/Instagram/WhatsApp, Google/G Suite/Gmail/YouTube, LinkedIn, Microsoft/M365, Signal, Slack, Twitter, and Wickr. Campaigns also pervasively used technology services focused on the political sector, including NGP Van/VoteBuilder, ActBlue, and WinRed. Securing accounts across so many providers took time. Differences across the various providers regarding the available security settings, and how those settings were named and used, increased the complexity. People involved with campaigns often did not have the time or technical knowledge to tackle this complexity, given that security was not among their top priorities.

\begin{displayquote}
``There's tons of <communication> channels now… between text message, GChat, Facebook, Wickr, Signal, WhatsApp, Slack, it's all of those.'' –A participant
\end{displayquote}

\paragraph{Shared accounts.}
People on campaigns usually maintained multiple shared email accounts (e.g., for the candidate, press, general information, donations, etc.) and social media accounts (e.g., Facebook, Twitter, Instagram, YouTube, etc.), to enable multiple staffers and consultants to communicate with voters, donors, and more. In some contexts, participants mentioned using email aliases like ``press@'' or ``info@'' to protect the privacy of workers who managed the alias, while providing an authoritative point of contact. Participants estimated that shared accounts might be accessed by 2 to 20 people. The shared nature of these accounts introduced security vulnerabilities: participants described them as often having passwords that were simple, credentials that were shared in potentially vulnerable ways (e.g., on a whiteboard, a note taped to a laptop, or via SMS or email), and 2FA not enabled, in part due to a lack of support for multi-tenant accounts.

\begin{displayquote}
``Not all services have a coherent model for shared accounts... Some like <Provider>, have no model. So there's no choice on <Provider>... around having a single set of account credentials that are shared by a large group of people. And that large group probably includes the candidate, random volunteers who walked in the door, and outside consulting firms. And so it's not shocking at all that these passwords are simplistic, and that these accounts get taken over with some regularity.'' –A participant
\end{displayquote}

\paragraph{Personal account use.}
Participants emphasized that campaign workers pervasively used personal (i.e., non-managed, consumer) accounts, for several reasons. First, they were incredibly busy, which led to ad hoc decision-making regarding what account to use in any given context. For example, if a staffer was on the go with only their personal device, they might use the personal accounts already set up on it. Relatedly, campaigns rarely had enough money to buy devices for staff or consultants, so personal device use was common. This meant that personal and work accounts were housed together on individual personal devices, and cross-over between the two was more likely in the busy and chaotic campaign work environment. Another barrier to managed account use was that sometimes providers charged money per account or for certain administrator features. Furthermore, in an effort to control who may be viewed as ``speaking on behalf'' of the campaign, some campaigns were hesitant to give consultants and volunteers campaign-domain email accounts or other managed accounts. Finally, participants explained that some consultants used consumer accounts (e.g., Gmail, Yahoo, Hotmail, or AOL) for campaign work. Even when participants described efforts to separate work and personal accounts, they still considered personal communications at risk. Personal accounts were a major vulnerability for campaigns, because they housed data that attackers were after, and campaigns had no visibility or control over their security configurations.

\begin{displayquote}
``What ends up happening is that work product ends up in their personal accounts. So <attackers> go after the personal accounts, where they have data on polling, on digital trends, plans, what emails we're going to send out... that sort of stuff. All of this is being shared in <cloud-based docs> to their personal account that IT staff can't secure.'' –A participant
\end{displayquote}

\paragraph{Transient account ownership.}
Incumbents are frequent campaigners, and many elected positions in the U.S. have no term limits. For example, in the U.S. Congress, people in the House of Representatives campaign every 2 years for their seat; Senators campaign every 6 years; some states' Governors and representatives campaign every 2 or 4 years. Campaigns that recurred tended to have new staff who reused existing accounts. Thus, a host of accounts needed to be passed from one owner (who may no longer be on the campaign) to the next, after each cycle. The need to pass on account ownership, frequently paired with a lack of knowledge of how to do this safely and efficiently, created a barrier to adoption for account security features, like strong passwords and 2FA. Participants also described cases where former staffers or consultants retained access to sensitive accounts.

\begin{displayquote}
``<After a campaign ends, you> usually give people 30 to 60 days to wind down, and then you'll keep 1 or 2 accounts around... Depending on whether the campaign is dead, or if it might come back in a couple of years… Then everything pretty much lies dormant. There's no real hygiene of going through and changing all the passwords. People leave me on <social media accounts> from campaigns for a long time... <Then if> they decide to run and reactivate their old website, their social media, that kind of thing, they may clean up who was on it, or they may not.... What happens most frequently is, <someone asks> ‘Hey do you remember the password to this one? I can't remember it, or I can't find it, or I don't have the recovery email.' I've dealt with that with every campaign I've ever worked on.'' –A participant
\end{displayquote}

\paragraph{Under-utilization of strong 2FA.}
Given their experiences with campaigns, participants believed that most people involved with campaigns had likely heard of 2FA, used some form of it on at least one account, and associated it with being an important part of account security. However, 2FA was described as under-utilized across campaign workers' many accounts. Personal accounts were commonly described as not protected by 2FA, even though they often contained campaign-relevant information or communications.

\begin{displayquote}
[Do you have 2FA on most accounts?] ``Not all of them, but some of them.'' [How do you decide which accounts?] ``Mostly, what I'm prompted for, what is called to my attention.'' –A participant
\end{displayquote}

People involved with campaigns were described as often using weaker second factors (SMS codes were commonly used). All but two participants told us what second factors they used—either at present, or in the past. Many used multiple types—including SMS codes, app-generated codes, codes from an email or phone call, security keys, hardware token codes, and/or prompts—but SMS codes were by far the most commonly used. All the security key users had some sort of background or training in computer security. Participants who had successfully helped colleagues adopt 2FA described needing to start them on more familiar phone-based factors and ease them into stronger second factors (which were widely perceived as harder to use).

\begin{displayquote}
``<I use phone-based prompts.> I've considered going to the super-advanced, key chain thing...'' [Why did yo\textit{}u decide not to use a hardware security key?] ``It seemed like it was a little more than I was bargaining for, from an ease-of-use standpoint… I felt like my level of security was sufficient… <The hardware security key is> another device.'' –A participant
\end{displayquote}

Our interview protocol included a set of questions to evaluate participants' understanding of 2FA and various second factors. Based on their responses, we inferred from participants that many of them and their campaign colleagues did understand that 2FA is for account security and has to do with signing in, but did not understand how 2FA improves their security, that different types of second factors offer meaningfully different levels of protection, and that they should use 2FA to protect most of their accounts.

\begin{displayquote}
 [What attacks might the code via phone call be vulnerable to?] ``I don't know… I have no idea.'' [How about the app?] ``It just seems really complicated to set up. But hacking I guess? I don't know.'' –A participant 
\end{displayquote}

Participants cited several pain points with 2FA that contributed to this low adoption or understanding. Most commonly cited was that 2FA required extra time and effort. This mattered to some busy participants, who described prioritizing their effort by only setting up 2FA on accounts perceived to be at risk (which left some personal accounts unprotected). Second, quick access to their accounts was described as critical, and some participants worried that 2FA might keep them out of accounts. For example, some traveled often and had trouble accessing second factors on airplanes, when they did not have cell service or WiFi, or when their phone's battery was dead. Others avoided using a security key because they worried about losing it or not having it when needed. Some cited the need to buy a physical thing as a barrier, either for cost or convenience reasons. Further, because most campaigns did not have full time IT staff, they did not have access to tech support for help with 2FA setup or account lockouts.

\begin{displayquote}
``This is important: inconveniencing candidates is one of the third rails of working on a campaign. And so, if you introduce any sort of wrinkle that might prevent the candidate from accessing their <social media> account, or being able to receive their email because they don't have their two-factor device there... that's a recipe for a bad day. Because they're going to get very upset… none of them thinks they're going to be targeted for cybersecurity attacks.'' –A participant
\end{displayquote}

Account lockout is a real risk with 2FA~\cite{peri2019login}. However, it seemed that the population overestimated the likelihood or frequency of this happening, and underestimated the importance of using stronger account security protections given the threats they face.

Finally, participants who were more familiar with security technologies noted that 2FA suffered from usability issues, caused in part by a lack of standardization across providers on terminology, requirements, available second factor options, and where settings were found.

\begin{displayquote}
``The reality is the industry has done a terrible job of making it easy to be secure by default. Even very basic things, like if people hear about two-factor and actually want to do it, good luck finding the settings. Good luck understanding why every service has different requirements... Everyone calls it something different...'' –A participant
\end{displayquote}

\paragraph{Weak password practices.}
Participants who had worked on party committees or more permanent organizations or consultancies, which were not transient like campaigns, typically described password practices that followed best practices, including common use of password managers, within their organizations. However, they observed that campaigns typically did not have such policies and relied more heavily on personal accounts. Thus, campaigns depended on individual workers choosing to use strong, unique passwords on their own personal accounts. And as noted above, passwords were often simplistic and sometimes shared insecurely for shared accounts, and were sometimes known by people no longer on the campaign for transiently owned accounts.

Password managers were not commonly used on campaigns, according to participants. Several participants with security expertise used password managers and had tried to help others on campaigns adopt them. They described these efforts as challenging: password managers incorrectly saved credentials often enough that confused and busy users abandoned them.

\begin{displayquote}
``<Campaign workers> should probably use a password manager. But… password managers are just way too awkward and complicated for people to bother with. And even when people really care, the reality is that the integration with the apps and the browsers is flaky enough that even when... you think you saved the password, but you really saved a different password, or you didn't actually save your password, and you never wrote it down or anything. So the next time you come to the site, whatever is in the password manager doesn't work. That happens often enough for me, and I'm committed to them. For people who aren't committed to them, that sort of thing happens once, they get confused once, and they're done.'' –A participant
\end{displayquote}

\subsection{Campaign threat models}
The practices and vulnerabilities described above are meaningful in the context of the specific threats people involved with campaigns believe they face. While prior work has detailed threats to campaigns~\cite{evanina2020,huntley_how_2020,burt_new_2020}, here we focus on what participants perceived the main threats to be.

\paragraph{Attackers.}
Participants were most concerned about nation-state attackers. According to participants, these attackers are sophisticated, well-funded, relentless, and had not yet experienced repercussions from the widely publicized attacks during the 2016 election cycle. Nearly all participants raised those attacks as evidence that some people involved with campaigns were targeted by nation-states (though not all believed they specifically were likely targets). A few participants described evidence that they and their colleagues had been targeted, including that they had received specialized state-sponsored attack warnings, or general security alerts on their accounts.

\begin{displayquote}
``It seems narcissistic in some ways to think that you're going to be the target of it, but I got over that because as recently as last month I received the warning that ‘a nation-state has attempted to access your account'… And I've gotten account recovery attempts, you know, ‘Someone is attempting to recover your account. Is this you?' And I'm like ‘no.''' –A participant
\end{displayquote}

Some participants reported concerns about thieves, citizens, or special interest groups as potential attackers. Most (but not all) participants were less concerned with the other party, members of the same party, or the press as attackers—but there was nearly universal concern about those entities getting leaked sensitive information as a result of an attack.

\paragraph{Attacks \& targeting.}
The attacks that participants had experienced, or heard about others in politics experiencing from colleagues or the media, were top of mind. Collectively, their top concern was phishing—they noted that it was cheap, easy, can lead to considerable damage, and worked in the 2016 election cycle. Our more security-knowledgeable participants noted that personal accounts, especially for communications and social media, were a focus of attackers since they housed data of interest to attackers and were less likely protected by an IT team, security expert, or security policies (e.g., that enforces 2FA). Multiple participants reported receiving phishing emails, though a few did not believe their role with campaigns increased their risk of receiving phishing messages.

\begin{displayquote}
``The <attacks> that I'm the most nervous about are phishing attempts that are getting more and more sophisticated… I've seen a lot of them... In this last 6 months or so... I've seen some really effective phishing attempts… The domain is spoofed really effectively... Those make me nervous because there are people on political campaigns who… it's not that they’re careless, it's just that they don’t know any better… They will click on things and have no idea what they might be opening up.'' –A participant
\end{displayquote}

Attackers were described as motivated to identify people affiliated with a campaign. Participants talked about how attackers might use FEC filings (which are publicly available, and list anyone paid by a campaign) and social media (staffers often update their employment status, listing the campaign). As evidence of this occurring, a participant recounted how people affiliated with a campaign—even across organizations—received the same phishing emails.

\begin{displayquote}
``Political organizations have to report to the FEC where they spend money, and who they pay, including staff and consultants. So we tell people: ‘You work <on a campaign> now. Your name will be on an FEC report. You will be targeted. ... Don't put it on <social media> right away. You're making yourself a target.''' –A participant
\end{displayquote}

Though less salient than phishing, participants discussed other attacks. From sophisticated attackers, participants talked about ransomware, malware, and DDoS. From domestic attackers, participants talked about people stealing devices or information from campaign offices, constantly following and recording candidates or campaign staff in person (to capture something damaging), harassing candidates or others involved with campaigns on social media, and threatening physical harm—all of which were experienced across our set of participants. From a variety of attackers, participants discussed scams aimed to steal money or efforts to spread mis/disinformation and fake news.

\paragraph{Harms.}
For campaigns, the main harm participants worried about was losing an election due to leaked or stolen information, or monetary theft. Leaked information could cause embarrassment, loss of voters, or the loss of a strategic advantage. For democracy, the harm was that election outcomes could be changed, weakening the institution of elections.

\begin{displayquote}
``As somebody who experienced it firsthand working on the 2016 race, I think that played a huge role in the outcome, based on the events that happened afterwards. If campaigns aren't secure, elections may not be free, and that's not what we stand for. If somebody can have a hand in what gets said based on using material that was hacked, I think that's very dangerous.'' –A participant
\end{displayquote}
\section{Recommendations for safer campaigns}
Political campaigns have recently realized the impact that digital security can have on candidates, staff members, political parties, and even democracy. The high profile attacks of 2016 were a catalyst, raising questions about the digital security responsibilities of campaigns. What we likely observed in our study is a population in the midst of a shift. We don't yet know the scope of the changes that campaigns will implement over the next decade. Time will reveal how their work culture evolves, what security norms become commonplace, and what tools and processes they adapt from security best practices (e.g., risk assessments, auditing, automated threat scanning, managed hardware, zero-trust environments, or hiring dedicated security professionals). Behavior change is difficult and often takes time. We believe that changing the direction of campaign security culture will require the help of a diverse group of experts, including from political committees and organizations, technology companies, and academia. 

In this section, we offer ideas to help guide the shift toward prioritizing security. In particular, we propose three near-term areas of investment for those who support campaigns: (1) establishing effective, consistent security guidance; (2) improving the usability of account security; and (3) improving the affordability of security protections. We also suggest long-term areas for future research. See Table 2 for a summary of our findings and recommendations.


\subsection{Education \& consistent guidance}
Most users do not prioritize security~\cite{herley2009,whitten1999}, and we saw a similar attitude among most campaign workers. We believe that security training for people involved with campaigns will be an important part of a work culture shift towards prioritizing security, and many of the experts involved in our roundtable agreed \cite{roundtable-report}.

Some participants described efforts by various party committees and organizations that are already underway to train campaign staff, consultants, and party committee members on security. For example, the DDC ramped up educational efforts across many federal campaigns in the 2020 election cycle and gave free hardware security keys to staff on over 140 campaigns. However, these efforts are relatively new, and they often reference guides (such as the D3P Cybersecurity Campaign Playbook~\cite{belfer_cybersecurity_2017}, Device and Account Security Checklist 2.0~\cite{lord_device_2019}, and others) that, while offering good security advice, differ in the order of which security actions they ask campaigns to prioritize. This echoes prior work on security advice for general Internet users, which shows that the security expert community lacks consensus~\cite{reeder_152_2017}. Inconsistent advice—even when the inconsistencies seem small to security experts—can cause confusion and inaction. Similar to Herley's recommendation to clearly prioritize security advice for users \cite{herley2009}, it would be helpful if the people and organizations that campaigns trust (political influencers, political committees, technology providers, policy makers, etc.) recommend the same top-priority security protections for campaigns to employ. People on campaigns would also benefit from consistent technical language when guidance from various sources refers to technical concepts and features.

Education and training efforts will be a key component of improving the security of campaigns going forward, though alone are not a panacea. To be successful, people should be trained as early as possible in their work with any campaign. Security guidance needs to be championed by the people who influence campaign work culture. Experts in our roundtable emphasized that this would typically include the campaign manager, senior staff, and certain consultants \cite{roundtable-report}. These are important people to train first. The candidate was not typically described as the “right” person to set security priorities, since they tend to focus on public appearances and fundraising, but many thought that candidates could be influential in communicating priorities for the campaign. We acknowledge that even with more consistent guidance, educational efforts will still be limited by the pace, resources, attention, and priorities of campaign workers.

\subsection{Usable account security}

While campaigns face a multitude of vulnerabilities and potential threats, we argue that solutions should prioritize the ability and willingness of people involved with campaigns to successfully use strong account security protections.

\begin{table*}[ht]
    \parbox{.37\linewidth}{
    \begin{tabular}{p{6cm}}
        \textbf{\large Perspectives from Campaigns}  \\
        \midrule
        \textbf{Work culture} \\
        -- Top priority: Win the election \\
        -- Transient \\
        -- Busy \& tight budgets \\
        -- Amorphous boundaries \\
        -- Limited security knowledge \\
        -- Security is not prioritized \\

        \midrule
        \textbf{Security practices \& vulnerabilities} \\
        \hspace{0.27cm}\textit{Communications} \\
        -- Sensitive \& targeted \\
        -- Many providers are used \\
        -- Ad hoc practices \\
        -- Busy workers overshare \\
        -- Campaigns do not control protections \\
        -- Encryption \& cloud use likely growing \\
        \hspace{0.27cm}\textit{Accounts} \\
        -- Many accounts across multiple providers \\
        -- Shared \& transiently owned \\
        -- Personal account use \\
        -- 2FA under-used \& misunderstood \\
        -- Weak password practices \\

        \midrule
        \textbf{Top perceived threats} \\
        -- Targeted attacks \\
        -- Phishing \\
        -- Nation-states \\
        \bottomrule
    \end{tabular}
    }
\hfill
    \parbox{.62\linewidth}{
    \centering
    \begin{tabular}{p{9.8cm}}        
        \textbf{\large Recommendations for Safer Campaigns} \\
        \midrule
        \textbf{Overall} \\
        -- Campaign security—including changing work cultures to prioritize security—must be a joint effort, involving technology companies, policy makers, political organizations, academic institutions, and campaigns \\[25pt]
        \midrule
        \textbf{Education} \\
        -- All supporters should provide consistent guidance on top security priorities for campaigns \\
        -- Train campaign workers as early as possible \\
        -- Include motivational content aimed to establish security culture on campaigns \\
        -- Recruit influencers (e.g., campaign manager, senior staff, consultants) to champion security \\
        \midrule
        \textbf{Technology} \\
        -- Standardize 2FA across providers \\ 
        -- Reduce password manager errors and overhead \\ 
        -- Evaluate usability of security products with campaign workers in situ \\
        -- Improve multi-tenant account usability \\
        \midrule
        \textbf{Policy} \\
        -- Ease financial burden on campaigns for security costs \\
        \midrule
        \textbf{Research community} \\
        -- More research on campaign security around the world, more types of campaigns/roles \\
        \bottomrule
         
    \end{tabular}
    }
    \caption{Summary of the themes in our findings and recommendations presented in this paper.}
\end{table*}

\paragraph{2FA.}
Usability issues with 2FA are known for creating adoption challenges, even in cases where the motivation to employ it appears to be high (e.g., for banking \cite{gunson20112FA,krol20152FA}). Similarly, our results show that 2FA adoption continued to lag among campaign workers, and even when present, weak options like SMS codes were among the most popular. Many participants had a poor understanding of 2FA, the protections offered by different second factors, and the importance of using 2FA on most accounts. We also learned that people involved with campaigns used many accounts and providers, but that the user experience of 2FA differed across providers. 

To better support people involved with campaigns, technology providers could explore how 2FA usability across different platforms can be improved—users are likely to be confused by differing terminology, requirements, second factor options, and locations of settings. Part of this includes exploring if it is possible to offer consistency in what second factors and other requirements providers support. And as part of the educational efforts described above, this population needs clear advice about what type of 2FA they should turn on and for what accounts, as well as help to understand why 2FA is important to winning elections.

\paragraph{Password practices.}
Passwords are the first layer of defense for accounts, and password managers can be an effective way to help users manage strong, unique passwords and minimize the chance that they will accidentally enter their authentication credentials into a phishing site. For shared accounts, password managers can help manage and secure strong credentials when multiple people need access. However, most participants thought that password managers took significant enough overhead to set up and use that widespread adoption by people involved with campaigns was very unlikely. 

Technology providers can better support this population by continuing to improve the integration and interoperability of password managers, browsers, and apps. Password managers could ensure users are explicitly warned when they attempt to enter credentials on an unknown site or one that might be a phishing site, with advice on how to tell if there is a problem.

\paragraph{Shared \& transiently owned accounts.}
Participants observed that shared and transiently owned accounts tended to have weak security protections. Technology providers can help by improving multi-tenant account usability and making it easier to audit and change account access and ownership.

\subsection{Policy: Affordable technology \& training}

Campaigns need help with the cost of security technologies and training. Regarding training, it is hard for campaigns to invest in training staff when they will be on to their next job in a year or less. Part of the return on investment with training is in the longer careers of those individuals in campaigns and politics, which is an expense that might be difficult for a budget-constrained campaign to justify. The typical staffer will change jobs frequently, and consultants work across many campaigns, so staffers and consultants who do receive security training can bring the knowledge and practices with them to the next campaign. Regarding security technologies, it is hard to justify investments in infrastructure that will no longer be used after election day. The decision to invest in security technology and training is easier if such resources are freely available. In many instances, campaign finance laws and regulations prevent companies and organizations from making contributions to campaigns, including providing free security services or technologies (such as hardware security keys or shared password manager accounts that cost money for others), absent specific exemptions~\cite{fec2018}. Policy makers could consider ways to enable all campaigns to leverage free, effective, easy-to-use security support, to protect themselves and democracy.

\subsection{Future research}

We believe that more research would help a diverse group of experts from political committees and organizations, technology companies, and academia improve security for political campaigns. They would benefit from foundational research exploring campaign security around the world, and with an even broader range of campaigns and campaign workers, including down-ballot races, candidates' family members, and more. Technology providers would benefit from usability studies of specific protections (such as 2FA, password management, shared accounts, and more), especially recruiting from this population and observing product use in realistic contexts.

\section{Conclusion}

\begin{displayquote}
“Security and politics should be separate. If you’re a candidate, you should win or lose on your best day, based on who you are. Not because your email got popped and posted online by a <nation-state>. –A participant
\end{displayquote}

Our study highlighted how the ongoing security challenges facing political campaigns stem from a combination of work culture, technology practices, and underdeveloped threat models. Campaigns are transient organizations with tight budgets and amorphous boundaries that are made up of busy people with limited security knowledge. Participants described digital security as a relatively new need for campaigns—one not often viewed as essential to winning elections—making investments in security hard to justify. 

People on campaigns presently rely on a variety of personal and work accounts across platforms and domains. Their ad hoc adoption of 2FA, strong passwords, encryption, and access controls introduced vulnerabilities that were not consistently mitigated. Participants recognized a growing risk of state-sponsored attacks (and phishing in particular), though expressed that strong protections continue to lag in adoption.

No one company, organization, institution, or campaign can solve the problems described in this paper on their own. Protecting campaign accounts and data will be more successful as a joint effort, involving a variety of perspectives and collective action from technology companies, the policy community, committees and organizations that support campaigns, academic institutions, and individual users who are involved with campaigns. We provide an initial understanding of this complex problem space that we hope will be used to help work toward solutions that are effective for this population. We suggest that, in the near-term, effective, consistent security guidance should be prioritized to inform security education; investigations should be performed on how to coordinate the standardization of usable account security protections (including 2FA, password managers, and multi-tenant accounts); and the affordability and availability of security technologies and training should be improved. In the longer-term, people working to support campaigns could explore how to shift the work culture of campaigns to prioritize security. With such collective action, we in the security community can do our part to improve digital security for campaigns, helping to protect future elections and democracy.

\section{Acknowledgments}

We thank everyone who participated in our research; all of our roundtable attendees and their assistants; our many colleagues at Google who helped make the research and roundtable happen; our paper reviewers; and our paper shepherd.

{
\footnotesize
\balance
\bibliographystyle{abbrv}
\bibliography{SPGP}
}

\section*{Appendix} For supplementary material related to our interview script, consent form, and codebook, please see
the ancillary file.
\end{document}